\documentclass{article}

\newcommand{\beq}{\begin{equation}}
\newcommand{\eeq}{\end{equation}}
\newcommand{\beqn}{\begin{eqnarray}}
\newcommand{\eeqn}[1]{\label{#1}\end{eqnarray}}

\newcommand{\id}
{i\kern.06em\hbox{\raise.25ex\hbox{$/$}\kern-.60em$\partial$}}

\newcommand{\bs}{/\kern-.52em b}
\newcommand{\qs}{/\kern-.52em s}

\newcommand{\dd}
{\kern.06em\hbox{\raise.25ex\hbox{$/$}\kern-.60em$\partial$}}

\newcommand{\tr}{\mathop{\rm tr}\nolimits}

\begin{document}
\begin{titlepage}

\title{Renormalization of the Hamiltonian and a geometric
interpretation of asymptotic freedom.}

\author{
G. Alexanian$^a$\thanks{E-mail address:
garnik@scisun.sci.ccny.cuny.edu}\,\,\,\,\, and \,\,\,\,\,
E. F. Moreno$^{a,b,c}$\thanks{Investigador CONICET, E-mail address:
moreno@scisun.sci.ccny.cuny.edu}\\
\\
~
\\
{\small\it $^a$ Physics Department,
City College of the City University of New York}\\
{\small\it New York NY 10031, USA}\\
~\\
{\small\it $^b$ Departamento de F\'\i sica,
Universidad Nacional de La Plata}\\
{\small\it C.C. 67, 1900 La Plata, Argentina}\\
~\\
{\small\it $^c$ Baruch College,
City University of New York}\\
{\small\it New York NY 10010, USA}\\
}


\maketitle

\begin{abstract}
Using a novel approach to renormalization in the Hamiltonian
formalism, we study the connection between asymptotic freedom and
the renormalization group flow of the configuration space metric.
It is argued that in asymptotically free theories the effective
distance between configuration decreases as high momentum modes
are integrated out.
\end{abstract}
\end{titlepage}


\section{Introduction}

Looking at the great success that the Standard Model has had
since it was introduced more than 20 years ago, it is quite
striking that we still lack understanding of the strong
interaction part in the low energy regime. In spite of excellent
results in the perturbative QCD we are unable to produce any
analytical computation of quantities such as the magnetic moment
of proton which is known with great accuracy. Some important
puzzles, like where does $\Lambda_{QCD}$ come from, still need to
be addressed. Apart from the numerous lattice results (which
reinforce our belief that QCD is the right theory for the strong
interactions) the situation remains largely the same  more than
20 years later. Many attempts to apply a variational approach
have not yet produced any effective calculational method for the
solution of the problems mentioned before.

It was suggested in \cite{feynman} that perhaps we need an
alternative way to look at the Yang-Mills theory based on a more
geometrical point of view. Namely, one tries to study quantum
mechanics of the fields in the space of gauge-inequivalent
configurations. The following analogy with quantum mechanics is
used. Consider a free particle inside a box of size $L$. The
lowest eigenvalue of the Hamiltonian (which is just a Laplacian)
is of the order of $\sim {1\over L^2}$ This is realized by the
state of the longest possible wavelength, $\lambda \sim  L$. It
is clear therefore, that the spectrum of such a system will have
a gap due to the fact that $L$ is finite. In other words, the
spectrum is going to be discrete as long as distances in the
configuration space cannot become arbitrarily large. Feynman's
suggestion was to try to adapt this idea to Yang-Mills theory.

In general, in order to determine the distance between field
configurations we need to know the metric of the configuration
space. The geometry of the configuration space of Non-Abelian
gauge theories was considered by many authors
(\cite{atiyah,Singer,Viale}) and, recently, in  \cite{orland}.
The major problem is to extract the metric on a space of the
gauge-inequivalent configurations ${\cal A}/{\cal G}$.

A natural distance between two arbitrary gauge configurations
$A_1$ and $A_2$ (that is gauge connections {\it modulo} gauge
transformations) is given by \cite{orland}
\beq
{||A_1-A_2||}^2=inf(\int_x Tr((A_1^{g}(x)-A_2(x))^2)|_g
\eeq
where $A^{g}_1$ is a gauge-transformed $A_1$. However the
extremum solution of this expression is highly non-local and very
difficult to work with. Nevertheless one can try to see some
qualitative features of this distance. In \cite{feynman} it was
argued that due to the non-Abelian nature of the gauge group the
low-potential-energy configurations are in a space of finite
diameter for 2+1 dimensional Yang-Mills theory. According to our
very crude analogy with the quantum mechanics it would mean that
the kinetic energy operator will have a discrete spectrum. Of
course, these are only qualitative arguments that require a
rigorous mathematical formulation if one is to draw any
conclusion on the existence and value of the mass gap. Also, as a
word of caution we must say that some properties of the
configuration space obtained in 2+1 dimensions may not
necessarily be the same as in the 3+1 dimensional case (as, for
example, the statement above in the case of YM in 3+1 and 1+1
dimensional sigma model \cite{orland,orland2}).

Another reason to believe that this approach could lead us to a
better understanding is the recent progress in Hamiltonian
formulation of the 2+1 dimensional Yang-Mills theory. In a series
of papers \cite{Nair1,Nair2} it was shown that by introducing
special gauge-invariant variables one can prove that the
(properly regularized) volume of the configuration space for the
non-Abelian theory is finite, while the corresponding quantity
for the Abelian field is infinite. The discreteness of the
spectrum of the kinetic term, $E^2$, has been also shown
explicitly and the string tension computed in \cite{Nair2} is in
remarkable agreement with the recent Monte-Carlo simulations
\cite{MC}.)

Behind all this discussion a natural question arises: What is the
behavior of the configuration space as we integrate out high
momenta degrees of freedom? Of course we can not answer
completely this question, and even a partial response deserves a
profound, probably non-perturbative, analysis. However we still
can say something, and answer specific matters as the behavior of
the configuration space {\it metric} under the renormalization
group, which encodes much of the properties of the space. Notice
that those issues acquire a special significance in Hamiltonian
formalism: there the kinetic term is essentially a Laplacian in
configuration space and the energy spectrum, at least in the
strong coupling regime, is dominated by it.

The choice of this topic was not fortuitous but we have an idea
in mind. In the standard understanding of asymptotic freedom all
the significance is put in the interaction potential through the
statement that the couplings decrease to zero as the energy at
which the theory is tested increases to infinity. However in the
Hamiltonian picture we can state the problem in a different way.
In the Hamiltonian there is an obvious competition between the
kinetic energy and the potential energy. So it is natural to
analyze the asymptotic freedom in its ``dual" form, {\it i.e.,}
the variation of the kinetic energy (with respect to potential
energy) as the renormalization scale changes. In particular we
conjectured the following, {\it in asymptotically free theories
the effective distance between configurations {\bf decreases} as
the momentum is lowered.}

Notice that this question does only make sense in a Hamiltonian
formalism, since it is only in this case that all the geometrical
properties of the configuration space can be precisely defined
(the Hamiltonian is an operator that acts on the Hilbert space of
functionals defined on the configuration space). It can be argued
that one can always work in a Lagrangian formalism and at the end
construct the respective Hamiltonian by the standard Legendre
transformation. However this {\it cannot} be done if
renormalization is involved at any moment. In the Lagrangian
formalism, even in the framework of perturbation theory, the
process of renormalization generates high time-derivative terms
in the Lagrangian, making impossible even the very definition of
a Hamiltonian, as the system is governed by high order
time-derivatives equations of motion. Even an application of the
Ostrogradsky method can in general lead to negative norm states
if higher time derivatives are involved in the Lagrangian. Thus,
a prescription for renormalization within the Hamiltonian
formalism is indispensable. One must not leave the phase space, thereby
maintaining the first order time derivative nature of the
equations of the system.

Therefore we will use a novel procedure of renormalization of
Hamiltonians, already introduced in ref. \cite{hrg} where a
successful renormalization of the Hamiltonian of $\lambda\;
\phi^4$ was performed. This method relies on the successive
diagonalization of the Hamiltonian by performing iterative
unitary transformations and subsequent projection onto the
Hilbert space of low energy modes. It resembles, in spirit, the
renormalization approaches of Glazek and Wilson \cite{GlWi},
Wegner \cite{We} and several other authors \cite{frohlich,
JoPeGl, HaOk, NM}, though in practice appears very different. To
support our conjecture we have done a detailed analysis of two
particular examples: Quantum Electrodynamics and Yang-Mills
theory in 3+1 dimensions, where we constructed the renormalized
Hamiltonian up to one loop.

The paper is organized as follows: In Section 2 we present
briefly our renormalization group technique for Hamiltonians. In
Sections 3, 4 and 5 we compute the 1-loop renormalization of the
$SU(N)$ Yang-Mills Hamiltonian. In Section 6 we did the same with
Quantum Electrodynamics. In Section 7 we discuss the relation
between renormalization of the metric and asymptotic freedom.
Finally, Section 8 contains some conclusions.


\section{Renormalization in the Hamiltonian formalism}

Let us review in this section the formalism of Hamiltonian
renormalization introduced in ref. \cite{hrg}.

Consider a system described by Hamiltonian $H$ which has already
some large cut-off $\Lambda$ built into it; that is the system is
ultraviolet-finite from the very beginning. We are assuming that
the Hamiltonian is written in terms of renormalized fields and
couplings up to the scale $\Lambda$ and incorporates all the
renormalization $Z(\Lambda)$ factors.

Now if $\mu$ is some intermediate scale, $\mu < \Lambda$, we want
to find the effective Hamiltonian $H(\mu)$ that has the same low
energy spectrum as the original Hamiltonian. More precisely, we
want to find the operator that has the same spectrum as the
original Hamiltonian when projected onto the Hilbert subspace
generated by the excitations with frequency less than $\mu$:
\begin{equation}
H(\mu) = {\cal P}_{low}\; H(\Lambda)\; {\cal P}_{low} .
\label{e1}
\end{equation}
where ${\cal P}_{low}$ is the projector onto the low frequency
subspace.

We will show that, in the framework of perturbation theory, it is
possible to partially diagonalize the Hamiltonian $H(\Lambda)$
and construct the vacuum state for the high frequency modes.
Thus, the low energy effective Hamiltonian would take the form

\begin{equation}
H(\mu) = \langle 0_{high} |U^{\dagger}(\Lambda,\mu)\;
           H(\Lambda)\;U(\Lambda,\mu)| 0_{high}\rangle.
\label{e2}
\end{equation}

Notice that it is not necessary to diagonalize completely the
Hamiltonian as we only need to identify the low energy states and
not the whole spectrum.

Suppose that after a unitary transformation we bring a
Hamiltonian to the form
\begin{equation}
H_{}({\rm low})+H_{\rm free}({\rm high}) + V
\label{e3}
\end{equation}
where $H_{}({\rm low})$ contains only low frequency
operators, $H_{\rm free}({\rm high})$ is a free Hamiltonian for
the high frequency modes and $V$ has the special form
\begin{equation}
V=\sum_{k,p>\mu}\;a^{\dagger}_k\; S_{k p}(\mu,a^{\dagger},a)\; a_p
\label{e4}
\end{equation}
with $S_{kp}(\mu,a^{\dagger},a)$ an arbitrary operator of low and
high frequency modes. Thus, using standard Rayleigh-Schroedinger
perturbation theory it follows that the correction to an
arbitrary state $|n\rangle$ is given by
\begin{equation}
|\delta n\rangle = \sum_{l\neq n} {
\langle l|V|n\rangle \over (E_l-E_n)}|l\rangle +
\sum_{l,n\neq m}{\langle l|V|m\rangle
\langle m|V|n\rangle \over (E_l-E_m)(E_m-E_n)}|l\rangle +\cdots
\label{e5}
\end{equation}
Therefore, if $|n\rangle$ is the vacuum state of $H_{\rm
free}({\rm high})$, it is annihilated by $V$ and there is no
correction to it at any order in perturbation theory.

Now we will show how to find the unitary transformation
$U(\mu,\Lambda)$ that transforms the Hamiltonian into the form
(\ref{e3}). First we split the Hamiltonian in four pieces:
\begin{equation}
H=H_1+H_2+V_A + V_B.
\label{e6}
\end{equation}
Here $H_1$ contains only the modes with $k<\mu$, $H_2$ is the
{\it free} part for all the modes with $k>\mu$, $V_A$ contains
the ``pure'' terms that have only high frequency creation
operators or high frequency annihilation operators, but not both,
and $V_B$ the remaining terms containing at least one
annihilation {\it and} one creation operator of high energy
modes(we assume here that $V_A$ and $V_B$ are normal-ordered with
respect to the free high frequency vacuum). Notice that the term
$V_B$ is already in the form (\ref{e4}). Then, according to the
preceding discussion, the objective of the renormalization
procedure is to ``eliminate" the terms in $V_A$ that spoils the
form (\ref{e3}, \ref{e4}).

We will proceed iteratively:  we write the unitary operator in
the form $U=U_0 U_1 U_2\cdots$, and compute each $U_n$
perturbatively. Each factor $U_n$ will bring the Hamiltonian to
the form (\ref{e3}) at a given order in $\mu/\Lambda$. Therefore,
there will be two expansion parameters in this procedure: one is
the coupling constant of the theory $\lambda$ and  the other is
the ratio $\mu / \Lambda$ (effectively
$\omega_{low}/\omega_{high}$).

Let us parametrize $U_n$ as $ exp(i\Omega_n)$, where $\Omega_n$
is a hermitian operator to be determined. We start from equation
(\ref{e6}) and perform a first unitary transformation expanding
in powers of $\Omega$ (in the general case $\Omega$ is at least
of order $\lambda$ so at a given order only a finite number of
terms are needed):
\begin{eqnarray}
e^{-i\Omega_0}(H_1+H_2+V_A + V_B)e^{i\Omega_0}&\!=&\!
H_1+H_2+V_A + V_B +
i\left[ H_1,\Omega_0\right]+ \nonumber\\
&&\hspace{-1cm} +\, i\left[ H_2,\Omega_0\right]+
i\left[V_A,\Omega_0\right]+i\left[V_B,\Omega_0\right]\cdots
\label{e7}
\end{eqnarray}

We want to eliminate the term $V_{A}$ which is of the first order
in $\lambda$, so $\Omega_0$ has to be of the same order.
Furthermore we want to generate an expansion in $\mu/\Lambda$, so
we impose the following condition on $\Omega_0$
\begin{equation}
i\left[ H_2,\Omega_0\right]+V_{A}=0.
\label{e8}
\end{equation}
This equation can be solved perturbatively and since commutators
with $H_2$ generate time derivatives we have the desired
expansion.

Notice that equation (\ref{e5}) defines $\Omega_0$  up to the
terms that commute with $H_2$. As it is shown in the appendix A
this freedom corresponds to the arbitrariness of the definition
of the low energy Hamiltonian up to a unitary transformation.
Therefore we will assume some kind of ``minimal" scheme - namely
that $\Omega_0$, after normal ordering, does not have a part that
commutes with $H_2$ (terms that are functions of $a^{\dagger}_k a_k$).

After $\Omega_0$ is chosen to cancel $V_{12}$ in the effective
Hamiltonian a new mixing term of order $\lambda$ has appeared
from $i\left[ H_1,\Omega_0\right]$. Notice however that this new
term is of higher order in $\mu/\Lambda$ and will be eliminated
by the next unitary transformation $U_1 = e^{i\Omega_1}$.
Explicitly,
\begin{eqnarray}
e^{-i\Omega_1}e^{-i\Omega_0}(H_1+H_2+V_A + V_B)
e^{i\Omega_0}e^{i\Omega_1} &\!\!\!=&\!\!\! H_1+H_2 +V_B
+\nonumber\\ &&\hspace{-7cm} i\left[ H_1,\Omega_0\right]+
 i\left[ H_2,\Omega_1\right]+
i\left[H_1,\Omega_1\right]+ {\rm second\;order\; terms}\cdots
\label{e9}
\end{eqnarray}
and we now choose $\Omega_1$ so that
\begin{equation}
i\left[ H_1,\Omega_0\right]+i\left[ H_2,\Omega_1\right] =0.
\label{e10}
\end{equation}
Using equations (\ref{e9}) and (\ref{e10}) we obtain:
\begin{eqnarray}
e^{-i\Omega_1}e^{-i\Omega_0}(H_1+H_2+V_{12})
e^{i\Omega_0}e^{i\Omega_1} &\!\!\!=&\!\!\! H_1+H_2 + V_B +
{i\over 2}\left[V_A,\Omega_0\right]+
\nonumber\\
&&\hspace{-6cm}i\left[V_B,\Omega_0\right]+
i\left[V_B,\Omega_1\right]
+ i\left[H_1,\Omega_1\right]-
{1\over 2}\left[\left[H_1,\Omega_0\right],\Omega_0\right]
-\nonumber\\
&&\hspace{-6cm}
{1\over 2}\left[\left[H_1,\Omega_0\right],\Omega_1\right]
-{1\over 2}\left[\left[H_1,\Omega_1\right],\Omega_1\right]+
\; {\rm higher\;order\; terms}\cdots
\label{e11}
\end{eqnarray}
The obvious next step is to introduce $\Omega_2$ in order to
cancel $i\left[H_1,\Omega_1\right]$ and continue with the same
process. Then a simple question emerges: Where should we stop? To
answer this question we have to consider the divergence
properties of the terms introduced by each new $\Omega$ into the
$H_{eff}$. These contributions like, for example, $-{1\over
2}\left[\left[H_1,\Omega_0\right],\Omega_0\right]$ may diverge as
$\Lambda \rightarrow\infty$. Nevertheless, the degree of
divergence  of each new term will be smaller as we introduce more
and more $\Omega_n$ factors. In general, the following power
counting can be used: since $\Omega_n$ is determined from
\begin{equation}
\left[ H_2,\Omega_{n+1}\right]=-\left[ H_1,\Omega_n\right]
\label{e12}
\end{equation}
and $H_2\approx\Lambda, H_1\approx\mu$, then
\begin{equation}
\Omega_{n+1}\approx{\mu\over\Lambda}\Omega_n .
\label{e13}
\end{equation}
Therefore the next $\Omega$ will be less divergent and eventually
all the new terms introduced by this prescription will be
convergent starting from certain $n$. At this point we will stop
since for our purposes we are only interested in divergent
contributions.

Of course so far we have only eliminated the high momentum degrees
of freedom up to the first order in the coupling constant.
Requiring the absence of the $\lambda^2$ - order mixing terms
will lead to the introduction of a whole new series of unitary
transformations, and the same arguments can be applied to them.

Finally, the actual process of renormalization is performed by
choosing the renormalization $Z(\Lambda)$ factors of the original
Hamiltonian to cancel the divergent contributions coming from
evaluation of (\ref{e2}). We now turn to the explicit computation
for QED and QCD Hamiltonians.


\section{$SU(N)$ Yang Mills theory: Preliminaries}

The kinetic energy term for the Hamiltonian of Yang-Mills theory
is essentially a Laplace operator on the configuration space.
However, due to the gauge invariance, the physical degrees of
freedom belong to the space of gauge connections modulo the group
of gauge transformations, {\it i.e,} the space of non-equivalent
gauge potentials, and  a satisfactory parametrization of this
space is needed. In this section we find a perturbatively
adequate coordinate system of the configuration space (a similar
analysis was done in ref. \cite{GS}) and compute the associated
metric .

We consider a $SU(N)$ Yang-Mills theory in the temporal gauge,
$A_0=0$. The canonical variables are the vector potential
$A_i^a(x)$ and the electric field $E^{a i}(x)$. They satisfy
canonical commutation relations
\beq
[A_i^a(x),E^{b j}(y)] = i \delta^{a b} \delta_i^j
\delta^{(3)}(x-y)
\label{m0}
\eeq
which permit the representation of the electric field as
\beq
E^{a i}(x)= - i \frac{\delta}{\delta A_i^a(x)}.
\label{m0.1}
\eeq
The Hilbert space of the theory is supplemented by the Gauss law
that enforces a constraint on the wave functionals, essentially
only allowing gauge invariant configurations,
\beq
{\cal G}^a(x) \Psi[A] = -i D[A]^{a b}_i\ \frac{\delta}{\delta A^b_i(x)}
\Psi[A]=0
\label{m0.2}
\eeq
where $D[A]_i^{a b} = \nabla_i \delta^{a b} - e f^{a b c} A_i^c(x)$
is the covariant derivative.

The gauge potential can be written also as a Lie algebra valued
field $A_i(x)= t^a A_i^a(x)$ where $t^a$ are the hermitian
generators of the Lie algebra of $SU(N)$ in the fundamental
representation, normalized to $\tr(t^a t^b)=\frac{1}{2} \delta^{a
b}$.

The Hamiltonian can be written as
\beq
H=\frac{1}{2}\int d^3 x \left( E^{a i}(x) E^{a i}(x) + B^{a i}(x)
B^{a i}(x)\right)
\label{m0.3}
\eeq
where we used the magnetic field $B_i^a$:
\beq
B^{a i}(x)=\epsilon_{i j k} \left(\partial_j A^a_k(x)+ \frac{1}{2} e
f^{a b c} A^b_j(x) A^c_k(x)\right).
\label{m0.4}
\eeq

Note that due to the constraint (\ref{m0.2}), not all the degrees
of freedom in the Hamiltonian (\ref{m0.3}) are ``physical". In
order to isolate the physical degrees of freedom we will perform
a change of coordinates in such a way that the Gauss law takes
its simplest form. We parametrize an arbitrary configuration in
terms of a ``gauge fixed" configuration plus gauge
transformations. The former will define a coordinate system of
the orbit space. We write
\beq
A_i(x) = g^{-1}(x) {{\cal A}_i(x)} g(x) + i g^{-1}(x) \partial_i g(x)
\label{m1}
\eeq
where $g$ is a $SU(N)$-valued matrix and ${\cal A}_i$ is a
configuration satisfying the Coulomb gauge condition
\beq
\nabla_i {\cal A}_i=0.
\label{m2}
\eeq
It is well known that the parametrization (\ref{m1}) with
condition (\ref{m2}) is not well defined globally due to the
Gribov ambiguity problem. However, we will work in the framework
of perturbation theory where the parametrization
(\ref{m1}-\ref{m2}) defines an acceptable isomorphism in the
configuration space.

{}From eq.(\ref{m1}) we deduce,
\beq
\delta A_i = g^{-1} \left( \delta{\cal A}_i + [D[{\cal A}]_i,
i \delta g g^{-1}] \right) g
\label{m3}
\eeq
where $D[{\cal A}_i] = \partial_i - i e {\cal A}_i$.

Thus we can decompose the metric of vector valued configurations
in``gauge fixed" and ``pure gauge" parts:
\begin{eqnarray}
\delta s^2 &=& \int d^dx\ \delta A_i^a(x)\ \delta
A_i^a(x)\nonumber\\ &=& \int d^3x\ \left(\delta {\cal A}_i^a(x)\
\delta {\cal A}_i^a(x) + D[{\cal A}]^{a b}_i (i \delta g g^{-1})^b
D[{\cal A}]^{a c}_i (i \delta g g^{-1})^c\right).
\label{m4}
\end{eqnarray}
The normalization of the wave functionals is then given by
\begin{eqnarray}
\langle \Psi_1|\Psi_2\rangle&=&\frac{1}{{\it Vol}\ G}
\int [DA_i^a]\ \Psi^*_1[A]\ \Psi_2[A]\nonumber\\
&=&\int [D{\cal A}_i^a]\ \sqrt{G}\ \Psi_1^*[{\cal A}]\
\Psi_2[{\cal A}]
\label{m4.1}
\end{eqnarray}
where $G$ is the matrix metric defined by equation (\ref{m4}).

Using the Coulomb gauge condition eq.(\ref{m2}) we can invert
eq.(\ref{m3}) and write
\beq
i\left(\delta g g^{-1}\right)^a=\left({\vec \nabla}\cdot {\vec D}[{\cal
A}]
\right)^{-1 a}_{b} {\vec \nabla} \cdot (\delta {\vec A} R^{-1})^b
\label{m5}
\eeq
and
\beq
\delta{\cal A}_i^a = \left( \delta_{i j} \delta ^{a b} -
D^{a c}_i[{\cal A}] \left({\vec \nabla}\cdot {\vec D}[{\cal A}]
\right)^{-1 c}_{a} \nabla_j \right)(\delta A_j R^{-1})^b.
\label{m6}
\eeq
where $R\equiv R(g)$ is the adjoint representation representative of
$g$.

Notice that not all the components of ${\cal A}_i$ are
independent as they are subject to condition (\ref{m2}), so we
can parametrize the space of gauge configurations modulo gauge
transformations with the $2 (N^2-1)$ functions ${\cal A}_{\hat
\imath}^a,\ \ {\hat \imath}=1,2$.

{}From eqs.(\ref{m5}-\ref{m6}) we can write the functional derivative as
\begin{eqnarray}
\frac{\delta \Psi}{\delta A_i^a(x)}& &=
\int R^{-1 ad} \left(R^{-1}_y \nabla_i^x({\vec D}
\cdot{\vec \nabla})^{-1}_{(y,x)}\right)^{b d}
\frac{\delta \Psi}{i(g^{-1}\delta g)^b(y)}+\nonumber\\
&&\hspace{-2cm}\int R^{-1 ad }\left(
\delta_{j i} \delta^{d b} \delta(x,y)
+ \left(D[{\cal A}]_j^{y}\nabla_i^x ({\vec D}\cdot{\vec
\nabla})^{-1}_{(y,x)}\right)^{b d}\right) P_{j {\hat \imath}}(y,z)
\frac{\delta \Psi}{\delta {\cal A}_{\hat \imath}^b(z)}
\label{m6.0}
\end{eqnarray}
where
$P_{i j}(x,y)=\delta_{i j}\delta(x,y) - \nabla_i\nabla_j\
\triangle^{-1}(x,y)$
is the projector on the transverse modes.

As previously announced, in these variables the Gauss law has the
simple form
\beq
\frac{\delta\, \Psi[A]}{(i\delta g\ g^{-1})^b} =0
\label{m6.01}
\eeq
so it is trivially imposed by demanding that the wave functionals
be independent of $g$.

It is useful to define a ``transverse" functional derivative as
\beq
\frac{\delta^T}{\delta {\cal A}^a_i(x)} = \int d^3z\ P_{i {\hat
\jmath}}(x,z)
\frac{\delta}{\delta {\cal A}^a_{\hat \jmath}(z)}
\label{m6.1}
\eeq
restoring rotational invariance at the expense of modifying the
canonical commutation relations:
\beq
\left[{\cal A}^a_i(x),\frac{\delta^T}{\delta {\cal A}^b_j(y)}\right]=
- P_{i j}(x,y)\ \delta^{a b}.
\label{m6.2}
\eeq

Now we can write the kinetic energy term in terms of the variables
${\cal A}^a_i$ (taking into account eq.(\ref{m6.1})),
\begin{eqnarray}
\langle \Psi_1|T|\Psi_2\rangle &=& \frac{1}{{\it Vol}\ G}
\int [DA_i^a]\ \frac{1}{2}
\int d^3 x \frac{\delta \Psi_1^*}{\delta A_i^a(x)}
\frac{\delta \Psi_2}{\delta A_i^a(x)}\nonumber\\
&=& \int [D{\cal A}_i^a]\ \sqrt{{\rm det}
G}\ \ \frac{1}{2}\int d^3 x\ d^3 y\ d^3 z \times\nonumber\\
&&\hspace*{-2.5cm}\left(\delta^{i j} \delta ^{a b}
\delta(x,y)
+ \left(D[{\cal A}]_j^{y}\nabla_i^x ({\vec D}\cdot{\vec
\nabla})^{-1}_{(y,x)}\right)^{b a}\right)
\frac{\delta^T \Psi^*_1}{\delta
{\cal A}^b_j(y)}\ \times\nonumber\\
&&\hspace*{-2.5cm} \left(\delta^{i k} \delta ^{a c}
\delta(x,z)
+ \left(D[{\cal A}]_k^{z}\nabla_i^x ({\vec D}\cdot{\vec
\nabla})^{-1}_{(z,x)}\right)^{c a}\right) \frac{\delta^T \Psi_2}
{\delta {\cal A}^c_k(z)}.
\label{m7}
\end{eqnarray}

That is, the kinetic energy density term takes the form
\beq
{\cal T} = -\frac{1}{2 \sqrt{G}} {\cal E}^{i a}(x) \sqrt{G}\
G^{(ia)(jb)}_{{\cal A}/{\cal G}}(x,y)\ {\cal E}^{j b}(y)
\label{m8}
\eeq
where ${\cal E}^{i a}=-i \frac{\delta^T}{\delta {\cal A}_i^a}$
and $G_{{\cal A}/{\cal G}}$ is the effective metric of the space
of gauge configurations module gauge transformations,
\begin{eqnarray}
G^{(ia)(jb)}_{{\cal A}/{\cal G}}(x,y)&=&\int d^3z\ \left(\delta^{i k}
\delta ^{a
c}
\delta(x,z) + \left(D[{\cal A}]_i^ {x}\nabla_k^z ({\vec D}\cdot{\vec
\nabla})^{-1}_{(x,z)}\right)^{a c}\right) \times\nonumber\\
&&\hspace*{-1cm} \left(\delta^{j k} \delta ^{b c}
\delta(y,z) + \left(D[{\cal A}]_j^{y}\nabla_k^z ({\vec D}\cdot{\vec
\nabla})^{-1}_{(y,z)}\right)^{b c}\right).
\label{m9}
\end{eqnarray}

As we mentioned above, the previous analysis was only valid in
the framework of perturbation theory where condition (\ref{m2})
defines a local system of coordinates on the orbit space. So it
is consistent with this approach to compute all the elements of
${\cal T}$ (the metric of the space of gauge transformations  $G$
and the metric of the true configuration space ${G}_{{\cal
A}/{\cal G}}$) in powers of the coupling constant $e$. In fact,
after a straightforward computation, the kinetic energy, up to
order $e^2$, can be shown to be
\begin{eqnarray}
T&=&\frac{1}{2} \int d^3x d^3y\ G^{(i a)(j b)}_{{\cal A}/{\cal G}}(x,y)
{\cal E}^{ai}(x) {\cal E}^{b j}(y)\ -\hspace*{3cm} \nonumber\\
&& \hspace*{3cm} \frac{i}{2} e^2 c_A\ \triangle^{-1}(0)
\int d^3x\ {\cal A}^a_i {\cal E}^{a i}(x)
\label{m10}
\end{eqnarray}
where $c_A$ is the Casimir of $G$ in the adjoint representation,
\beq
G^{(i a)(j b)}_{{\cal A}/{\cal G}}(x,y)=\delta^{i j} \delta^{a b}
\delta(x,y) - e^2 f^{a c d} f^{b c e} {\cal A}^d_i(x)
{\cal A}^e_j(y) \triangle^{-1}(x,y) \ + O(e^3)
\label{m11}
\eeq
and the transverse variables ${\cal A}_i^a$ and ${\cal E}_i^a$
satisfy the commutation relations (see eq.(\ref{m6.2}):
\beq
[{\cal A}^a_i(x), {\cal E}_j^b(y)] = i \delta^{a b}\ P_{i j}(x,y).
\label{m12}
\eeq


\section{Renormalization of Yang-Mills Hamiltonian }

In this section we will compute the renormalization contribution
to the $E^2$ and $B^2$ terms in the effective Hamiltonian. In
what follows, we will assume the ``gauge-fixed'' variables from
the previous section and will use $A$ and $E$ instead of $\cal A$
and $\cal E$. Using expression (\ref{m10}) the gauge-fixed
Yang-Mills Hamiltonian can be written as follows:
\begin{equation}
H={1\over 2}\int_{x,y}\alpha^{ab}_{ij}(x,y)E^a_i(x)E^b_j(y)+
{1\over 2}ic_A A^a_i(x)E^a_i(x)G(x,x) + {1\over 2} B^a_i(x) B^a_i(x)
\label{q0}
\end{equation}
where $c_A$ is $N$ for $SU(N)$ and
\begin{eqnarray}
\alpha^{ab}_{ij}(x,y) & = & \delta_{ij}\delta^{ab}\delta^3(x-y)+
e^2 f^{adc}f^{bde}A^c_i(x)A^e_j(y)G(x,y)+ \cdots
\label{q1}
\\
G(x,y) & = & \int {1\over q^2}\ e^{-iq(x-y)}{d^3q\over (2\pi)^3}
\label{G}
\end{eqnarray}
This is the so called ``bare'' Hamiltonian. For the loop
calculations we have to introduce the $Z$-factors by the
following procedure. Using (\ref{q1}) let us rewrite expression
(\ref{q0}) as
\begin{eqnarray}
H={1\over 2} Z_{E^2}(\Lambda) E^2+
{1\over 2} Z_{AAEE}(\Lambda)e^2 f^{adc} f^{bde}
A^c_i(x)A^e_j(y)G(x,y)E^a_i(x)E^b_j(y) \nonumber  \\
+{1\over 2} Z_{B^2}(\Lambda) B^2 +
e Z_1(\Lambda) f^{abc}\partial_i A^a_j A^b_i A^c_j +
{e^2 \over 4} Z_4(\Lambda) f^{abc} f^{dec}
A^a_i A^b_j A^d_i A^e_j+\cdots
\label{H}
\end{eqnarray}
Each of the $Z$-factors will have the following form
\begin{equation}
Z=1 + e\,f_1(\Lambda)+e^2\,f_2(\Lambda)+\cdots
\end{equation}
The functions $f_n$ will be chosen order-by-order from the
requirement that after integration of the modes from $\mu$ to
$\Lambda$ all the corrections sum up in such a way that
$f_n(\Lambda) \rightarrow f_n(\mu)$ and $Z(\Lambda)\rightarrow
Z(\mu)$, accordingly. When doing the one-loop corrections one can
therefore assume that all the $Z$'s are initially 1 and choose
the corresponding $f$'s from the condition that high-cutoff
dependence be cancelled after computing the $H_{eff}$ to one
loop.
Since one-loop wave function renormalization in QCD is of the
second order in coupling constant it is easy to see that we need
$\Omega$ only up to the first order in $e$. Then there is only
one term of $V$ that is relevant:
\begin{equation}
V^{(3)}=e\, f^{abc}\partial_i A^a_j A^b_i A^c_j .
\label{q2}
\end{equation}
To compute the renormalization of the quadratic terms in the
Hamiltonian we have to assign, according to the general procedure
of Section 2, two A's to be ``high'' and one ``low''. Therefore,
the relevant part of $V^{(3)}$ looks like \footnote{We use the
following convention: fields with subscript $1$ indicate ``low
momenta" fields and fields with subscript $2$ indicate ``high
momenta" fields}
\begin{equation}
V^{(3)}=e\, f^{abc} ( 2\partial_i A^a_{1j}A^b_{2i}A^c_{2j}+
\partial_i A^a_{2j} A^b_{1i} A^c_{2j})
\label{q3}
\end{equation}
We now write $A_2$ and $E_2$ in second-quantized form,
\begin{eqnarray}
A^a_{2i}(x)=\sum_k{1\over {(2\omega_k)}^{1\over 2}}
\left(a^a_{ki} e^{ikx}+a^{a\dagger}_{ki} e^{-ikx}
\right) \label{q4} \\
E^a_{2i}(x)=\sum_k i {\left(\omega_k\over 2\right)}^{1\over 2}
\left(a^{a\dagger}_{ki} e^{-ikx} - a^a_{ki} e^{ikx}
\right) ,
\label{q5}
\end{eqnarray}
where the creation and annihilation operators satisfy
\begin{equation}
\left[a^a_{ki},a^{b\dagger}_{pj}\right]=
\left(\delta_{ij}-{k_ik_j\over k^2}\right)
\delta^{ab} \delta_{kp}
\label{q6}
\end{equation}
After normal-ordering, $V^{(3)}$ can have three kind of terms:
$a^\dagger_k a^\dagger_p$, $a_k a_p$ and $a^\dagger_k a_p$. As
was explained in Section 2, only the first two types will be used
to in order to determine $\Omega$, so the final form of $V^{(3)}$
is therefore
\begin{eqnarray}
V^{(3)}&\!\!\! =&\!\!\! e\ f^{a b c}\ \sum_{p,q}\int d^3x\
\left\{ 2\left(
a^{a\dagger}_{p i} a^{c\dagger}_{q j} e^{-i(p+q)x}+
a^a_{p i} a^c_{q j} e^{i(p+q)x}\right)
\partial_jA^b_i(x) -\right.\nonumber \\
&&\hspace{-1cm} i q_j\left. \left(
a^{a\dagger}_{p i} a^{b\dagger}_{q i} e^{-i(p+q)x} -
a^a_{p i} a^b_{q i} e^{i(p+q)x}\right)
A^c_j(x) \right\}\frac{1}{2 \sqrt{\omega_p\ \omega_q}}
\label{q7}
\end{eqnarray}
In order so solve equation (\ref{e8}) with the interaction
given by (\ref{q3}) we use
\begin{equation}
H_2=\sum_k \omega(k) a^{b\dagger}_{ki} a^b_{ki},\ \
[H_2,a^{a\dagger}_{pi}]= \omega_p a^{a\dagger}_{pi},\ \
[H_2,a^{a}_{pi}]=-\omega_p a^{a}_{pi}\, .
\end{equation}
Then,
\begin{eqnarray}
\Omega_0&\!\!\! =&\!\!\! i\ e\ f^{a b c}\ \sum_{p,q}\int d^3x\
\left\{ 2\left(
a^{a\dagger}_{p i} a^{c\dagger}_{q j} e^{-i(p+q)x}-
a^a_{p i} a^c_{q j} e^{i(p+q)x}\right)
\partial_jA^b_i(x) -\right.\nonumber \\
&&\hspace{-1cm} i q_j\left. \left(
a^{a\dagger}_{p i} a^{b\dagger}_{q i} e^{-i(p+q)x} +
a^a_{p i} a^b_{q i} e^{i(p+q)x}\right)
A^c_j(x) \right\}\frac{1}{2 \sqrt{\omega_p\ \omega_q}
\ (\omega_p + \omega_q)}.
\label{q8}
\end{eqnarray}
According to the equation (\ref{e10}), the next $\Omega$ will be
\begin{eqnarray}
\Omega_1&\!\!\!\! =&\!\!\!\!\!
-i\ e f^{a b c}\ \sum_{p,q}\int d^3x
\left\{ 2\left(
a^{a\dagger}_{p i} a^{c\dagger}_{q j} e^{-i(p+q)x}+
a^a_{p i} a^c_{q j} e^{i(p+q)x}\right)
[H_1,\partial_jA^b_i(x)] -\right.\nonumber \\
&&\hspace{-1cm} i q_j\left. \left(
a^{a\dagger}_{p i} a^{b\dagger}_{q i} e^{-i(p+q)x} -
a^a_{p i} a^b_{q i} e^{i(p+q)x}\right)
[H_1,A^c_j(x)] \right\}\frac{1}{2 \sqrt{\omega_p\ \omega_q}
\ (\omega_p + \omega_q)^2}.
\label{q9}
\end{eqnarray}

To study the renormalization of the metric we have to determine
the corrections to the $E^2$ term in the Hamiltonian. There are
only two possibilities. The first one is $-{1\over
2}\left[\left[H_1,\Omega_0\right],\Omega_1\right]$, since
$\Omega_0\sim A$ and $\Omega_1\sim E$. The other comes from the
normal-ordering of the quadratic part of the $\alpha$ term in the
equation (\ref{q1}). When computing the double-commutator, there
is only one divergent term,
\begin{eqnarray}
-{1\over 2}\left[\left[H_1,\Omega_0\right],\Omega_1\right] &=&e^2
\sum_{k,p}
{1\over 2} \left[H_1,A^b_i(x)\right]\left[H_1, A^n_\alpha(y)\right]
f^{abc}f^{mnl}k_i q_\alpha\,\times \nonumber\\
&&\hspace{-3cm}
\left(\delta^{am}_{kq}\delta^{cl}_{ps}+
\delta^{al}_{ks}\delta^{cm}_{pq}\right)\,
P_{j\gamma}(k)P_{j\gamma}(p)\, \left( e^{-i(k+p)(x-y)}
+ e^{i(k+p)(x-y)}\right)\,\times\nonumber\\
&&\hspace{-2.2cm}
{1\over 4\, (\omega_k +\omega_p) (\omega_q +\omega_s)^2
\sqrt{\omega_k \omega_p \omega_q \omega_s}} .
\end{eqnarray}
At this point we can say that momenta $(k+p)$ and $-(k+p)$ are
essentially the momenta of the ``low'' fields. Therefore
$|k+p|<\mu$. We can now change the summation by the following
trick: say $ k+ p = r$ and $|r|<\mu$. Then $k=r-p$ and the
summation goes over $k$ and $r$. Any possible divergence can only
come from the summation over $k$. Using
$f^{adc}f^{bdc}=c_A\delta^{ab}$ the expression (27) can be
simplified as follows,
\begin{equation}
\sum_{r,k}
{c_A\over 4\omega_k \omega_p}
{\left[H_1,A^c_i(r)\right]\left[H_1, A^c_\alpha(-r)\right]
\over (\omega_k +\omega_p)^3}
\left(k_i k_\alpha - k_i p_\alpha \right)P_{j\gamma}(k)P_{j\gamma}(p)
\end{equation}
Noticing that the leading divergence of the expression is
logarithmic, which means that we can neglect the difference
between $k$ and $p$ (for the divergent contribution only) and
using
\begin{eqnarray}
\sum_{k}=\int {d^3 k\over (2\pi)^3},\ \ \ \ \omega_k=|k|,\ \ \ \
k\sim -p\\
P_{j\gamma}(k)P_{j\gamma}(p)=
\left(1+{(k\cdot p)^2\over k^2 p^2}\right)
\sim 2 ,
\end{eqnarray}
we obtain
\begin{equation}
{c_A\over 24}
\sum_{r}\left[H_1,A^c_i(r)\right]\left[H_1, A^c_i(-r)\right]
\sum_k{1\over \omega^3_k}=
-{c_A\over 24}{1\over 2\pi^2} log\left({\Lambda\over\mu}\right)
\int d^3x\,E^2
\label{q10}
\end{equation}
As we mentioned earlier, there is another term that can correct
the $E^2$ term of the effective Hamiltonian. It is the ``Coulomb
interaction'' term from the kinetic term in (\ref{q1}). Looking
at the $e^2$ order correction part of the (\ref{q1},\ref{q2}) and
choosing the A's to be ``high'' and the E's to be ``low'' we have
\begin{equation}
{e^2\over 2} f^{adc}f^{bde}
\int_{x,y} A^c_{2i}(x)A^e_{2j}(y)G(x,y)E^a_{1i}(x)E^b_{1j}(y)
\label{q11}
\end{equation}
As written, this term is not normal-ordered with respect to the
high-energy vacuum. Using (\ref{q4}),(\ref{q5}) and (\ref{q6}) we
obtain
\begin{equation}
{e^2\over 2}  f^{adc}f^{bde}
\int_x\int_y\sum_k
{\delta^{ce}P_{ij}(k)\over 2\omega_k} e^{ik(x-y)}
G(x,y)E^a_{1i}(x)E^b_{1j}(y)
\label{q12}
\end{equation}
Upon using  the definition (\ref{G}) of $G(x,y)$ one can see that
the leading divergence is logarithmic and that the final
expression reads
\begin{equation}
{e^2\over 6}\, c_A \,
\sum_k{1\over\omega_k^3}\int d^3x\;E^2\,
= {e^2\over 6} {c_A\over 2\pi^2}\,
log\left({\Lambda\over\mu}\right) \,\int d^3x\;E^2.
\label{q13}
\end{equation}
The sum of the terms (\ref{q10}) and (\ref{q13}) gives the total
correction to the kinetic energy at one loop:
\begin{equation}
\delta H_{E^2}={c_A\over 8}{{e^2}\over 2\pi^2}log\left({\Lambda
\over\mu}\right)\, \int d^3x\, E_i^2(x)
\label{q14}
\end{equation}
At this point we can say that the $Z_{E^2}$ factor is therefore,
\begin{equation}
Z_{E^2}=1-{c_A\over 4}{{e^2}\over 2\pi^2}log\left({\Lambda\over
m}\right)+\cdots
\label{q15}
\end{equation}
Since the operator $E_i$ is represented by a variational
derivative with respect to gauge field, $\delta\over \delta A_i$,
one would naturally expect that $Z_{E^2}$ should be equal to the
$Z_{B^2}^{-1}$. It is therefore an important check on our method
to show that it is indeed so. To compute $Z_{B^2}=Z_3$ we need to
find out $B^2$ correction to the effective H. In comparison with
the computation of the $E^2$ correction it is much more involved
due to the fact that for most of the terms $B^2$ comes as a
sub-leading divergence. We will not present  detailed computation
but sketch  the main steps and give the final result. $B^2$
contributions can arise from the following terms: ${i\over
2}\left[V,\Omega_0\right]$ and normal ordering of the $\alpha$
term from the equation (\ref{q1}) again. This term is similar to
the (\ref{q11}), the only difference being how the ``high'' and
``low'' components are assigned:
\begin{equation}
{e^2\over 2} f^{adc}f^{bde}A^c_{1i}(x)A^e_{1j}(y)G(x,y)
E^a_{2i}(x)E^b_{2j}(y).
\label{q16}
\end{equation}
Leading divergence for both terms is quadratic and gives
correction of the form $A^2$. The appearance of this term is
related to our choice of the cut-off procedure as a way of
regulating the theory; it can be dealt with by introducing a
$A^2\Lambda^2$ counter-terms in the bare Hamiltonian and defining
appropriate boundary conditions at the ends of the RG flow
trajectory \cite{ERG}. To capture the logarithmic contribution
one has to expand the denominators of the $\Omega_0$ up to the
second order in the momenta of the ``low'' fields. Tedious but
straightforward  computation  gives
\begin{equation}
-{i\over 2}\left[V,\Omega_0\right] =
-{27 \over 120}\,e^2\,c_A \,\left( \sum_k
{1\over k^3} \right)  \int d^3x\, B^2 .
\label{q17}
\end{equation}
The logarithmic divergent part of the normal ordering of the two
$E$'s from (\ref{q16}) gives
\begin{equation}
{1\over 10}\,e^2\,c_A \,\left( \sum_k {1\over k^3} \right)
\int d^3x\, B^2
\label{q18}
\eeq
The final result is
\beq
\delta H_{B^2}=-{c_A\over 8}{{e^2}\over 2\pi^2}log\left({\Lambda
\over\mu}\right)  \int d^3x\,B^2
\label{q19}
\eeq
which makes the corresponding $Z$ factor
\begin{equation}
Z_{B^2} = Z_3 = 1+{c_A\over 4}{{e^2}\over 2\pi^2}log
\left({\Lambda\over m}\right)+\cdots
\label{q20}
\end{equation}
This coincides with the value of $Z_3$ for QCD in the Coulomb
gauge obtained in the Lagrangian formalism \cite{Bernstein}.
%


\section{Three point function renormalization}

Let us show briefly the renormalization of the three point
function within our formalism. To do that let us recall the
general method of Section 2. We will have to be a little more
careful in the analysis of the relevant contributions to the
renormalized Hamiltonian.

According to the notation of Section 2, we write the Yang-Mills
Hamiltonian in terms of the low momentum and high momentum fields
as:

\beq
H=H_1 + H_2 + (V^{(3)}_A + V^{(3)}_B + V^{(4)} + \cdots)
\label{3p1}
\eeq
where $H_1$ is the part of the Hamiltonian that only contains low
momentum fields, $H_2$ is the part that only contains high
momentum fields, and the $V$'s are the ``mixing" terms. For
convenience we have separated these last terms according to the
vertex number and the high momentum creation-annihilation
operators structure:
\begin{eqnarray}
V^{(3)}_A\!&=&\! e\ f^{a b c}\ \sum_{p,q}\int d^3x\ \left\{ 2\left(
a^a_{p i} a^c_{q j} e^{i(p+q)x} +
a^{a\dagger}_{p i} a^{c\dagger}_{q j} e^{-i(p+q)x}\right)
\partial_jA^b_i(x) +\right.\nonumber \\
&&\hspace{-1cm} i q_j\left. \left(
a^a_{p i} a^b_{q i} e^{i(p+q)x} -
a^{a\dagger}_{p i} a^{b\dagger}_{q i} e^{-i(p+q)x}\right)
A^c_j(x) \right\}\frac{1}{2 \sqrt{\omega_p\ \omega_q}}\ ,\\
V^{(3)}_B\!&=& \! e\ f^{a b c}\ \sum_{p,q}\int d^3x\ \left\{ 2\left(
a^{c\dagger}_{q j} a^a_{p i} e^{i(p-q)x} +
a^{a\dagger}_{p i} a^{c}_{q j} e^{-i(p-q)x}\right)
\partial_jA^b_i(x) +\right.\nonumber \\
&&\hspace{-1cm} i q_j\left. \left(
a^{a\dagger}_{p i} a^b_{q i} e^{-i(p-q)x} -
a^{b\dagger}_{q i} a^{a\dagger}_{p i} e^{i(p-q)x}\right)
A^c_j(x) \right\}\frac{1}{2 \sqrt{\omega_p\ \omega_q}}\ ,\\
V^{(4)}\!&=&\! \frac{e}{2} \sum_{p,q}\int d^3x \left\{
\left(a^a_{p i} a^c_{q i} e^{i(p+q)x} +
a^{a\dagger}_{p i} a^{c\dagger}_{q i} e^{-i(p+q)x}\right)
f^{a b e} f^{c d e} A^b_j(x) A^d_j(x)\right.\nonumber\\
&&\hspace{-1cm} +\left.\left(a^a_{p i} a^b_{q j} e^{i(p+q)x} +
a^{a\dagger}_{p i} a^{b\dagger}_{q j} e^{-i(p+q)x}\right)
\left(f^{a b e} f^{c d e} + f^{a d e} f^{c b e}\right)
A^c_i(x) A^d_j(x)\right\}\times \nonumber\\
&&\frac{1}{2 \sqrt{\omega_p\ \omega_q}}.
\label{3p2}
\end{eqnarray}
Now we have to analyze which terms contribute to the three point
vertices. At this order only $\Omega$ at order $e$ is needed,
moreover, one can convince oneself that only $\Omega_0$, {\it i.e,}
the first iteration of the unitary transformation leads  to
divergent contributions. Then eq.(\ref{e7}) reads in this case:
\begin{eqnarray}
H'&=&H_1 + H_2 + V^{(3)}_A + V^{(3)}_B + V^{(4)} + \cdots +
i[H_1,\Omega_0] + i[H_2,\Omega_0] +\nonumber\\
&&\hspace{-1cm} i[V^{(3)}_A,\Omega_0] + i[V^{(3)}_B,\Omega_0] +
i[V^{(4)},\Omega_0]-
\frac{1}{2}[[H_1,\Omega_0],\Omega_0]- \nonumber\\
&&\hspace{-1cm} \frac{1}{2}[[H_2,\Omega_0],\Omega_0] -
\frac{1}{2}[[V^{(3)}_A,\Omega_0],\Omega_0] -
\frac{1}{2}[[V^{(3)}_B,\Omega_0],\Omega_0] - \nonumber\\
&&\hspace{-1cm} \frac{1}{2}[[V^{(4)},\Omega_0],\Omega_0] -
\frac{i}{6}[[[H_1,\Omega_0],\Omega_0],\Omega_0]
- \frac{i}{6}[[[H_2,\Omega_0],\Omega_0],\Omega_0]
+ \cdots
\label{3p3}
\end{eqnarray}

As explained in Section 2, we choose $\Omega_0$ in in such a way
that its commutator with $H_2$ cancels the mixing terms that
contain high momentum annihilation operators or high momentum
creation operators, but not both. Also, we note that up to this
order, only $V^{(3)}$ is of order $e$. Then $\Omega_0$ satisfies:
\beq
[H_2,\Omega_0]=i V^{(3)}_A
\label{3p4}
\eeq
(which is precisely the equation that gives eq.(\ref{q8}) ).

Now it is not difficult to individualize the only terms that
contribute to the three point vertex:
\beq
\delta H^{(3)}= \Delta H_1 + i[H_1,\Omega_0]^{(3)} +
i[V^{(4)}_A,\Omega_0]^{(3)} -
\frac{1}{2}[[V^{(3)}_B,\Omega_0],\Omega_0]^{(3)}
\label{3p5}
\eeq
Here $\Delta H_1$ stands for the normal ordering contribution
(tadpole diagram) of order $e^3$ in the kinetic energy. In fact,
at this order, the kinetic energy has a term of the form:
\beq
T_{e^3} = e^3 f^{a b c} f^{a d e} f^{e m n} \int_{x y z}
A_i^b(x) G(x,y) A_j^d(y) \partial_j G(y,z) A_k^m(z)
E^c_i(x) E^n_k(z)
\label{3p6}
\eeq
which, when properly contracted, generates the contribution:
\beq
\Delta H_1=\frac{1}{6} c_{A}\ e^3\ log\left(\frac{\Lambda}{\mu}\right)
f^{a b c} \int_x \partial_i A_j^a A_i^b A_j^c.
\label{3p7}
\eeq

The remaining terms in eq.(\ref{3p5}), are computed similarly to
the ones of Section 2. After a very lengthy, but straightforward
computation, we get the following results:
\begin{eqnarray}
[H_1,\Omega_0]^{(3)}&=& - i\frac{5}{24} c_{A}\ e^3\
\frac{1}{2 \pi^2}
log\left(\frac{\Lambda}{\mu}\right) f^{a b c} \int_x \partial_i
A_j^a A_i^b A_j^c\ ,\nonumber\\
{}[V^{(4)},\Omega_0]^{(3)}&=& i \frac{3}{8}
c_{A}\ e^3\
\frac{1}{2 \pi^2}
log\left(\frac{\Lambda}{\mu}\right) f^{a b c} \int_x
\partial_i A_j^a A_i^b A_j^c\ ,\nonumber\\
{}[[V^{(3)}_B,\Omega_0],\Omega_0]^{(3)}&=&-
\frac{1}{6} c_{A}\ e^3\
\frac{1}{2 \pi^2}
log\left(\frac{\Lambda}{\mu}\right)
f^{a b c} \int_x \partial_i A_j^a A_i^b A_j^c
\label{3p8}
\end{eqnarray}

Finally, adding up all the contributions we have
\beq
\delta H^{(3)}=\frac{1}{12} c_{A}\ e^3\
\frac{1}{2 \pi^2}
log\left(\frac{\Lambda}{\mu}\right) f^{a b c} \int_x \partial_i
A_j^a A_i^b A_j^c.
\label{3p9}
\eeq
This result corresponds to a renormalization constant $Z_1$ equal
to:
\beq
Z_1=1-\frac{1}{12} c_{A}\ e^2\
\frac{1}{2 \pi^2} log\left(\frac{\Lambda}{m}\right).
\label{3p10}
\eeq
which is the same as the Coulomb gauge result in the Lagrangian
approach as well \cite{Bernstein}.


\section{QED}
In this section we will outline a similar computation for Quantum
Electrodynamics. The QED Hamiltonian can be written as follows
\begin{equation}
H={1\over 2} \int d^3x\,\left( E^2+  B^2\right)+
\int d^3x\, \left\{
\bar{\psi}\left(i{\vec \gamma}\cdot{\vec \partial}+m\right)\psi+
e\, \bar{\psi}A\cdot\gamma \psi \right\}.
\label{qed1}
\end{equation}
The imposition of the Gauss law constraint,
\begin{equation}
\partial_i E^i = e {\bar \psi} \gamma^0 \psi
\label{qed2}
\end{equation}
generates an ``instantaneous'' Coulomb interaction and $H$ takes
the form
\begin{eqnarray}
H&=&{1\over 2} \int d^3x\, \left( E^2+ B^2\right) +
{1\over 2}\int_x\int_y e^2\,\bar{\psi}(x)\gamma^0 \psi(x) G(x,y)
\bar{\psi}(y)\gamma^0\psi(y) \nonumber \\
&&+\int d^3x\, \left\{ \bar{\psi}
\left(i{\vec\gamma}\cdot{\vec\partial}
+m\right)\psi+
e\bar{\psi}A\cdot\gamma
\psi \right\}
\label{qed3}
\end{eqnarray}
where $G(x,y)$ is given by the (\ref{G}). According to the
general idea we are supposed to split it in to ``high'' and
``low'' energy parts,
\begin{eqnarray}
H_1 &=&  {1\over 2} \int d^3x\, \left( E_1^2+ B_1^2\right) +
\int d^3x\, \bar{\psi_1}
\left(i{\vec\partial}\cdot{\vec\gamma}+m\right)\psi_1
+e\bar{\psi_1}A\cdot\gamma\psi_1 \\ \nonumber
H_2 &=&  \sum_k\omega_k b_{k\alpha}^\dagger b_{k\alpha}+
\sum_p\omega_p d_{p\beta}^\dagger d_{p\beta}\\
V^{(3)}&=&e \int d^3x\, \left\{\bar{\psi_2}A_1\cdot\gamma\psi_2
+e\bar{\psi_2}A_2\cdot\gamma\psi_1+
e\bar{\psi_1}A_2\cdot\gamma\psi_2
\right\}
\\
V^{(4)}&=&\int_x\int_y e^2\bar{\psi_2}(x)
\gamma^0 \psi_1(x) G(x,y)\bar{\psi_1}(y)\gamma^0 \psi_2(y)+\cdots
\label{qed4}
\end{eqnarray}
Here dots mean that we show only those terms that play role in
one loop effects. Using arguments similar to those in Yang-Mills
theory one can see that in order to determine $Z$ factors up to
the order $e^2$ we need $\Omega$ only up to the first order. It
turns out that only first two iterations in $\mu/\Lambda$ are
needed - $\Omega_0$ and $\Omega_1$. In parallel to the previous
sections, we determine $Z_3$ and $Z_1$ by identifying corrections
of the type $E^2$, $B^2$ and $\bar{\psi}A\cdot\gamma\psi$. There
is only one commutator that can contribute to the $B^2$ term:
${i\over 2}[V,\Omega_0]$ where one has to take subleading
divergence to identify the $log$ corrections. $E^2$ correction is
given by two commutators $-{1\over2}[[H_1,\Omega_0],\Omega_1]$
and commutator of the ``Coulomb'' term with $\Omega_0$ :
$i[V^{(4)},\Omega_0]$. The final result is
\begin{eqnarray}
\delta H_{E^2}&=& -{e^2\over 6}{1\over 2\pi^2}
log\left(\Lambda\over\mu\right)\int d^3x\, E_1^2\\
\delta H_{B^2}&=& {e^2\over 6}{1\over 2\pi^2}
log\left(\Lambda\over\mu\right)\int d^3x\, B_1^2
\label{qed5}
\end{eqnarray}
which leads us to the well-known answer for the $Z_3$-factor in
QED:
\begin{eqnarray}
Z_{E^2}^{-1}=Z_{B^2}=Z_3= 1 -{e^2\over 3}{1\over 2\pi^2}
log\left(\Lambda\over m\right)
\label{qed6}
\end{eqnarray}
Up to this point all our $Z$'s were identical to those known from
the covariant calculations in the Coulomb gauge. However, when
computing the other two renormalization constants: $Z_1$ and
$Z_2$ we find results which are different from the covariant
ones. In the case of $Z_2$ - fermion kinetic term renormalization
there are three possible contributions arising from ${i\over
2}\left[V,\Omega_0\right]$, $-{1\over 2}\left[\left[H_1,\Omega_0
\right],\Omega_0\right]$ and normal-ordering of the four-fermion
term in the Hamiltonian (\ref{qed3}). Extracting
$\bar{\psi}{\vec\gamma}\cdot{\vec\partial}\psi$-type corrections
from each of these terms one can see that they cancel. The
similar thing happens in case  of $Z_1$ as well: all the
$\bar{\psi}\gamma\cdot A\psi$-type terms cancel at the $e^3$
order. This makes both $Z_1$ and $Z_2$ equal to 1 at one loop.
Nevertheless there is no contradiction between our result and the
conventional one. In covariant formalism there exists a Ward
Identity $Z_1=Z_2$ which is essential for the maintaining the
gauge invariance of the effective action. It is  satisfied,
presumably, in our case as well. But values of $Z_1$ and $Z_2$
are gauge dependent and cancel out from the final expression that
defines the beta function for QED, which is determined by $Z_3$
only. There is an obvious reason why, say $Z_2$ must be 1 in our
case. Since $\bar\psi$ and $\psi$ are  conjugated variables, in
the Hamiltonian formalism one should be represented by the
variational derivative with respect to the other. Then, similarly
to the $E^2$ and $B^2$ terms for the gauge field, they will have
inverse renormalization factors which will cancel each other in
the final expression for the kinetic term for fermions.


\section{Geometric interpretation of the asymptotic freedom.}

Let us recall the Yang-Mills Hamiltonian, after integrating the
high momenta modes down to the scale $\mu$ (at order $e^2$),
incorporating the correct renormalization factors $Z_i$.
\begin{eqnarray}
H_{YM}&=&{1\over 2}\, {1\over Z_3(\mu)} E^2 + {1\over 2}\, Z_3(\mu)
(\partial_i A^a_j - \partial_j A^a_j)^2 + Z_1(\mu)\, e_R f^{abc}
\partial_i A^a_j A^b_i A^c_j + \nonumber\\
&&Z_4(\mu)\, {1\over 4} e^2_R f^{abc} f^{dec} A^a_i A^b_j A^d_i A^e_j +
\cdots
\label{conc1}
\end{eqnarray}
Similar expression for the QED Hamiltonian will be:
\begin{equation}
H_{QED}={1\over 2}\, {1\over Z_3(\mu)} E^2 + {1\over 2}\, Z_3(\mu)
B^2 +
Z_2(\mu)\bar{\psi}\;i\;{\vec\gamma}\cdot{\vec\partial}\psi+
Z_1(\mu)\, e_R  \bar{\psi}A\cdot\gamma\psi \cdots
\label{conc1.1}
\end{equation}
Here $e_R$ is a fixed quantity at certain scale and all the
dependence on the scale is hidden in the $Z$ factors. The
definition of the ``renormalized" fields through the
incorporation of the $Z$ factors was done in analogy with the
Lagrangian covariant approach where the renormalized quantities
are included in such a way that the ``renormalized" effective
action gives finite results when the cut-off is removed. But in
the Schrodinger picture this requirement is not necessary as the
fields are only coordinates of the configuration space and do not
enter explicitly in the computation of correlation functions.
With this fact in mind we will alter this requisite and adapt it
to our needs.

The usual covariant renormalization program puts the emphasis on
the interactions, and the scaling properties of the theory  are
extracted from the study of the $\beta$-functions. However, in a
Hamiltonian description the kinetic term plays a significant role
since essentially it is nothing but the Laplacian in
configuration space. Hence, many of the properties of the QFT can
be inferred from the geometrical features of the configuration
space (as compactness, boundness, etc.). In particular we are
interested in the change of the configuration space metric under
the renormalization group flow. We claim that in asymptotically
free theories the distance between configurations increases as we
move  to the UV limit, thus ``flattening" the potential energy
and consequently fading the interaction. We will support our
claim with the analysis of the one-loop Yang-Mills theory.

It is clear then, that in the spirit of our work we want to
stress the kinetic term (better, the configuration space metric)
over the potential energy and try to understand the asymptotic
behavior of the theory through the renormalization flow
properties of the distance in the configuration space. For this
reason we will rescale the fields in such a way to transfer the
renormalization scaling properties to the kinetic term.

Then let us rescale the fields as:
\begin{equation}
A_i^a \to \frac{1}{e_R}\ Z_3(\mu) Z_1(\mu)^{-1}\  A_i^a\ ,
\ \ \ \ \ \ \ \ \
E_i^a \to  e_R\ Z_3^{-1}(\mu) Z_1(\mu)\ E_i^a
\label{conc2}
\end{equation}
so the QCD Hamiltonian takes the form:
\begin{eqnarray}
H_{QCD}&=&{1\over 2}\, e_R^2\, \left(\frac{Z_1^2}{Z_3^3}\right)\
{E_i^a}^2 +
\frac{1}{e_R^2}\, \left(\frac{Z_3^3}{Z_1^2}\right)\left\{{1\over 2}
(\partial_i A^a_j - \partial_j A^a_j)^2 + \right.\nonumber\\
&&\hspace{-.5cm} \left. + f^{abc} \partial_i A^a_j A^b_i A^c_j
+{1\over 4} f^{abc} f^{dec} A^a_i A^b_j A^d_i A^e_j
\right\}
+ \cdots
\label{conc3}
\end{eqnarray}
where we have used the Slavnov-Taylor identities (adapted for the
Hamiltonian formalism) \footnote{ By writing both the Hamiltonian
$H$ and Gauss law ${\cal G}^a$ in terms of renormalized fields
with the explicit $Z$ factors and requiring $[H,{\cal G}^a]=0$
one can get linear relations between $Z$'s that lead to
(\ref{conc4}).}
\begin{equation}
\frac{Z_1}{Z_3}=\frac{Z_4}{Z_1}.
\label{conc4}
\end{equation}
Note that with this normalization we have ``homogenized" the
potential term (up to an overall factor) by transferring all the
cut-off dependence to the kinetic term. Now we can read from the
kinetic term, the cut-off dependence of the (inverse of the)
metric. In fact, using the result of Section 3, we can write the
cut-off dependent configuration metric as:
\begin{equation}
G_{(ia)(jb)}(x,y;\mu)=\left(1+ e_R^2\, {11 N\over 12}\,
{1\over 2\pi^2}\ {\log}(\mu/m_R)\right)\, G^0_{(ia)(jb)}(x,y) +
{\cal O}(e_R^3)
\label{conc5}
\end{equation}
where $G^0_{(ia)(jb)}$ is the metric defined in equation
(\ref{m9}) written in terms of the rescaled fields (\ref{conc2})
and $e_R$.

Looking at the QED Hamiltonian and using $Z_1=1$ and $Z_2=1$ one
can see that no rescaling is needed at all, since all the cut-off
dependence is already shifted to the kinetic term,
\begin{equation}
H_{QED}={1\over 2}\left(\frac{1}{Z_3}\right)\
E^2 + {1\over 2} Z_3
B^2 +
\bar{\psi}\;i\;{\vec\gamma}\cdot{\vec\partial}\psi+
e_R  \bar{\psi}A\cdot\gamma\psi \cdots
\label{conc3.1}
\end{equation}
Following our procedure the corresponding metric will be
\begin{equation}
G_{(ij)}(x,y;\mu)=\left(1 - e_R^2\, {1\over 3}\,
{1\over 2\pi^2}\ {\log}(\mu/m_R)\right)\, \delta_{(ij)}\delta(x-y) +
{\cal O}(e_R^3) .
\label{conc5.1}
\end{equation}
At this point it is easy to compare relative behavior of the two
metrics under the renormalization flow. Equation (\ref{conc5})
clearly shows that the distance between configurations decreases
as the cut-off is lowered, while the corresponding expression
(\ref{conc5.1}) increases, sustaining then our claim.
Incidentally, it is worthwhile to mention that the combination
$\frac{Z_1^2}{Z_3^3}$ that appears in equation (\ref{conc3}) is
precisely the one that defines the $\beta$-function of the
Yang-Mills theory and, at least in the Lagrangian approach, it
has been proved to be independent of the gauge fixing condition.
Similarly, QED expression involves only $Z_3$, since $Z_1=Z_2$ is
1 to this order, and $Z_3$ is the only constant that determines
beta-function for QED and it is known to be gauge-independent to
all loops.


\section{Summary and Conclusion}

There are several issues that are more natural to address in the
Hamiltonian picture than in the usual covariant Lagrangian
formalism. One of them, which we are interested in, is the
relevance of the geometry of the configuration space to the
properties of the corresponding Quantum Field Theory. The reason
is simple: in the Hamiltonian formalism the kinetic energy term
is nothing but a Laplacian operator in the configuration space
and its topological and geometrical features determine the nature
of its spectrum. Then it is natural to ask what is the behavior
of the configuration space as we integrate out high momentum
degrees of freedom.

To answer a small part of this question was the aim of this
paper. To be precise we were interested in the following aspect
of the problem: the evolution of the distance between field
configurations (and more precisely the metric) with the
renormalization group in asymptotically free theories. In
particular we state the following conjecture: {\it in
asymptotically free theories the effective distance between
configurations decreases as high momenta degrees of freedom are
integrated out.}

To support this statement we first developed an original
renormalization group technique for Hamiltonian formalism in the
framework of perturbation theory. This method  resembles the
Hamiltonian renormalization approaches of Glazek and Wilson
\cite{GlWi} and Wegner \cite{We} and operates by a progressive
diagonalization of the Hamiltonian by means of a succession of
iterative unitary transformations followed by a projection onto
the Hilbert space of the low-momentum degrees of freedom. Finally
we applied the formalism to two conspicuous QFT's: Quantum
Electrodynamics and Yang-Mills theory in 3+1 dimensions, where we
constructed the renormalized Hamiltonian up to one loop.

Our results were substantially supportive of our conjecture. In
the case of Yang-Mills, an asymptotically free theory, the
one-loop metric renormalization showed that in fact the distance
between configuration increases as the momentum scale increases,
and on the contrary for QED, not asymptotically free, the
behavior of the metric is the opposite.

We are aware, of course, that our results are not decisive but
just consistent with the conjecture. After all we have only
studied two examples at one-loop order in perturbation theory.
However from the examples considered we can observe a pattern
that seems to repeat at any instance: when moving all the weight
of the renormalization group onto the configuration space metric,
it acquires a renormalization factor which is a function of the
same combination of renormalization constants that defines the
$\beta$-function of the theory, and thus, presumably,  inheriting
its asymptotic behavior properties.


\section*{Acknowledgements}

We want to thank  Prof. V.P. Nair for suggesting this problem and
his advice, encouragement and input during the work on this
project. We are also grateful to Prof. P. Orland for helpful
discussions and critical reading of the manuscript. G.A. was
partially supported by CUNY RF grant 6684591433. E.F.M. was
supported by CONICET and CUNY Collaborative Incentive Grant
991999.

\section*{Appendix A}


In this appendix we show that the ambiguity in the definition of
the effective Hamiltonian due to the freedom in the solution of
equation (\ref{e8}) (and generally (\ref{e12})) is just the
standard ambiguity of the Hamiltonian operator, namely the
freedom of unitary transformations.

Let us recall that the solution of (\ref{e8}), (\ref{e10}) or
(\ref{e12}) is not uniquely defined; if $\Omega_n$ is a solution
so will be $\Omega_n + O_n$ with $O_n$ satisfying the homogeneous
equation $[H_2,O_n]=0$. Now let us consider two sets of
$\Omega$'s - say $\Omega_n^{(a)}$ and $\Omega_n^{(b)}$, all of
them satisfying the proper equations (\ref{e5}) and (\ref{e9})
but generated from different type of solutions. Then we have
\begin{eqnarray}
H_A&=& \dots\; e^{-i\Omega_1^{(a)}}
e^{-i\Omega_0^{(a)}} H e^{i\Omega_0^{(a)}} e^{i\Omega_1^{(a)}}
\dots \nonumber\\
H_B&=& \dots\;  e^{-i\Omega_1^{(b)}}
e^{-i\Omega_0^{(b)}} H e^{i\Omega_0^{(b)}} e^{i\Omega_1^{(b)}}
\dots
\label{ap1}
\end{eqnarray}

Obviously $H_A$ and $H_B$ are unitarily related: $H_A= U^{-1}
H_B U$ and consequently they have identical spectrum. What we have
to show is that this property remains {\it after} projecting onto
the ``high'' perturbative vacuum $|0_{high}\rangle$.

Following the prescription given in section 2, we have shown
that, up to given order $n$ in coupling constant $\lambda$ and a
given order $m$ in $\mu/\Lambda$ we can write:
\begin{eqnarray}
H_{A}&=&H_1^{(a)}({\rm low})+ H_2^{\rm free}({\rm high})
+ \sum_{k,p>\mu} a^{\dagger}_k S^{(a)}_{kp} a_p +
O\left( \lambda^n,\left({\mu/\Lambda}\right)^m\right)\nonumber\\
H_{B}&=&H_1^{(b)}({\rm low})+ H_2^{\rm free}({\rm high})
+ \sum_{k,p>\mu} a^{\dagger}_k
S^{(b)}_{kp} a_p+
O\left( \lambda^n,\left({\mu/\Lambda}\right)^m\right)
\label{ap2}
\end{eqnarray}
where the Hamiltonians $H_1^{(a,b)}$ depend only on the low
energy modes, $H_2$ is the free Hamiltonian for the energy modes
and $S^{(a,b)}$ are {\it arbitrary} operators of low and high
frequency modes.

Without losing any generality we can assume that $H_1^{(a)})$ and
$H_1^{(b)}$ are diagonal, as they can always be brought to that
form with a ``low energy modes"-unitary transformation that
respects the structure of (\ref{ap2}).

Now we will show that the low energy Hamiltonians  $H_1^{(a)}$
and $H_1^{(b)}$ have the same spectrum and consequently they are
unitarily related.

Consider the eigenvalues equations for $H_1^{(a)}$ and
$H_1^{(b)}$:
\begin{equation}
H_1^{(a)}\;\psi^a_{\alpha}= E^{a}_{\alpha}\;  \psi^a_{\alpha}
\; , \hspace{1cm}
H_1^{(b)}\;\psi^b_{\alpha}= E^{b}_{\alpha}\;  \psi^b_{\alpha}
\label{ap3}
\end{equation}

Using standard perturbation theory we can compute the eigenvalues
of the whole Hamiltonians, $A$ and $B$ as an expansion in powers
of the matrix elements of the interaction terms
\begin{equation}
V^{(a,b)} = \sum_{k,p>\mu} a^{\dagger}_k S^{(a,b)}_{kp} a_p .
\label{ap4}
\end{equation}

We get, for the eigenvalues of the operator $H_A$,
\begin{equation}
E^{\rm TOT}_{\alpha,n} = E^a_{\alpha} + E^0_n + \langle \alpha,n|
V^a| \alpha, n \rangle + \sum_{\gamma,m}  \frac{\langle \alpha,n|
V^a|\gamma,m\rangle \langle \gamma, m |V^a|\alpha, n
\rangle}{E^a_{\gamma} + E^0_m - E^a_{\alpha} - E^0_n } + \cdots
\label{ap5}
\end{equation}
and a similar equation is valid for the eigenvalues of the
operator $H_B$. But the low energy spectrum corresponds to those
states with $n=0$, and in this case, due to the particular form of
the interaction, all the perturbative contributions vanish and
the eigenvalue $E^a_{\alpha}$ is the {\it exact} eigenvalue of
the whole Hamiltonian (up to the given order in $\lambda$ and
$\mu/\Lambda$):
\begin{equation}
E^{\rm TOT}_{\alpha, 0} = E^a_{\alpha} +
O\left( \lambda^n,\left({\mu/\Lambda}\right)^m\right)
\label{ap6}
\end{equation}

And finally, since both Hamiltonians $H_A$ and $H_B$ have the
same spectrum, or since $E^{TOT}$ is the same for both $H_A$ and
$H_B$, we deduce that
\begin{equation}
E^a_{\alpha} = E^b_{\alpha} +
O\left( \lambda^n,\left({\mu/\Lambda}\right)^m\right) \;.
\label{ap7}
\end{equation}
%


\end{document}